# Topological defects govern crack front motion and facet formation on broken surfaces

Itamar Kolvin, Gil Cohen and Jay Fineberg

*The Racah Institute of Physics, the Hebrew University of Jerusalem, Jerusalem, Israel 91000*

**Patterns on broken surfaces are well-known from everyday experience, but surprisingly, how and why they form are very much open questions. Well-defined facets are commonly observed[1-4] along fracture surfaces which are created by slow tensile cracks. As facets appear in amorphous materials[5-7], their formation does not reflect microscopic order. Fracture mechanics, however, predict that slow crack fronts should be straight, creating mirror-like surfaces[8-13]. In contrast, facet-forming fronts propagate simultaneously within different planes separated by steps. It is therefore unclear why steps are stable, what determines their path and how they couple to crack front dynamics. Here we show, by integrating *real-time* imaging of propagating crack fronts with surface measurements, that steps are topological defects of crack fronts; crack front separation into discontinuous overlapping segments provides the condition for step stability. Steps drift at a constant angle to the local front propagation direction and the increased local dissipation due to step formation couples to the long-range deformation of the surrounding crack fronts. Slow crack front dynamics are enslaved to changes in step heights and positions. These observations show how 3D topology couples to 2D fracture dynamics to provide a fundamental picture of how patterned surfaces are generated.**

The surface patterns created by cracks are objects of fascination as well as practical utility [1,4], but the fundamental laws that govern their formation remain obscure. Classically, cracks are treated as 2D; fracture surfaces are reduced to a line ending at a singular point – the crack tip – where the two surfaces are created. Cracks will propagate when the energetic cost of breaking a unit surface area, the fracture energy $\Gamma$, is balanced by the energy flow to the singular crack tip, $G$; $G = \Gamma$. Crack velocities are bounded by the Rayleigh surface wave speed, $c_R$. This framework[14] is very successful in predicting the dynamics of simple tensile cracks that produce structure-less "mirror" surfaces[15,16]. In 3D, the crack tip becomes a singular *line*, the *crack front*. The appearance of structure[1,4] and roughness[17] within fracture surfaces demonstrates the need for a 3D picture. Theory has, however, shown[8-13] that slow ($\ll c_R$) tensile (Mode I) crack fronts are stable to perturbations, although the addition of twist (Mode III) can destabilize crack fronts[18-22]. Surprisingly, experiments have shown that purely tensile cracks *can* spontaneously form structure; facetted fracture surfaces appear in brittle materials ranging from crystalline silicon to amorphous polymers[1-7]. Here we examine how these non-trivial facets are created and how they couple to crack front dynamics.

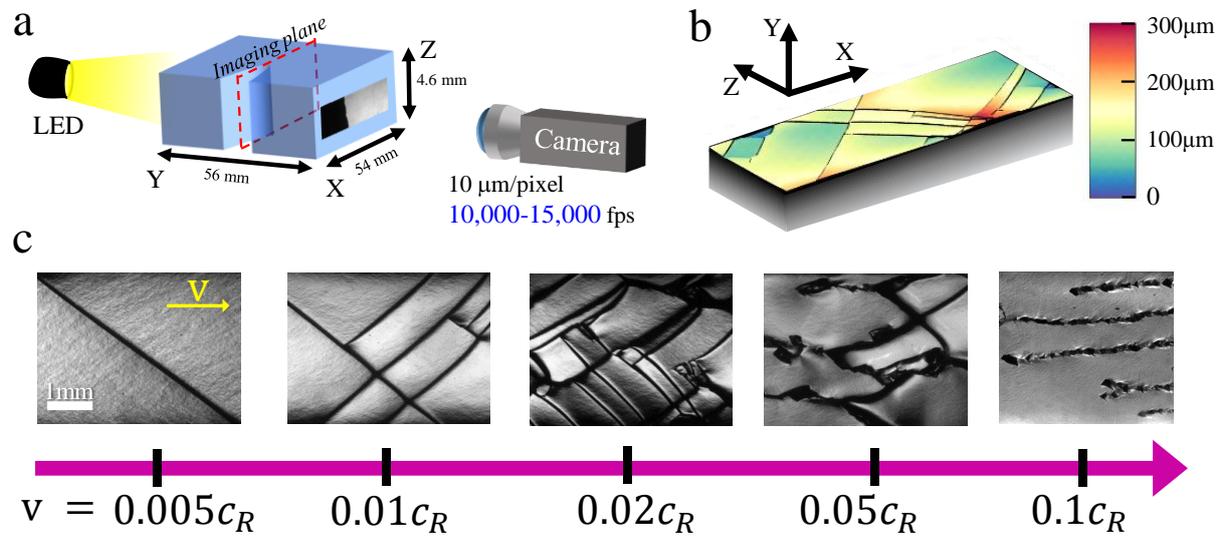

*Figure 1.* **Fracture surface patterns in brittle polyacrylamide gels**. (a) The experimental set-up. A rectangular gel slab is held in imposed tension along the Y axis. The sample is illuminated by a collimated LED beam in the Y direction. The light is deflected by the crack tip producing shadow images of the propagating front that are captured by a fast camera. (b) A profilometric surface scan performed after the experiment reveals a typical facetted surface. (c) Images of typical fracture surfaces formed by cracks with increasing velocities. For $v < 0.01c_R$ fracture surfaces are either mirror-like, or contain a single step-line (*left*). As v is increased, step nucleation becomes more frequent (*2nd and 3rd panels*). At $v \sim 0.05c_R$ micro-branches, which are localized kite-shaped structures, appear and coexist with step-lines (*4th panel*). For $v \gtrsim 0.1c_R$ step-lines disappear altogether and micro-branches appear in chains or branch-lines aligned parallel to the local front propagation direction (*right*). For our gels $c_R = 5.2 \ m/s$.

Our experiments are performed in brittle polyacrylamide gels of dimensions (X,Y,Z) = (54,56,4.6)mm under uniform tensile loading (Methods). In gels, crack speeds are reduced by 2-3 orders of magnitude[23] ($c_R \sim 5 \ m/s$) compared with hard brittle materials, enabling crack dynamics to be captured in unprecedented spatial and temporal resolution. We visualize crack fronts propagating along the X axis in real-time by shining light through the transparent sample along the tensile Y axis (Fig. 1(a)). The high curvature at the crack front deflects the light and forms a shadow image that is captured by a high-speed camera. We correlate crack front dynamics with the surface structure they generate via optical profilometer (Fig. 1(b)) measurements.

In Fig.1(c) we present typical structures formed on fracture surfaces. These are highly sensitive to mean front velocities, v. At $v < 0.01c_R$, surfaces are either mirror-like or consist of two facets separated by a step-line. Once nucleated, step heights grow and stabilize at $40 \pm 10 \mu m$. Steps then drift at an angle to the crack front. Nucleation is strongly facilitated near existing steps and sample boundaries. As crack velocities increase, step nucleation becomes increasingly frequent leading to intersections and mergers of step-lines and increased velocity fluctuations. When v ~ $0.05c_R$, micro-branches, localized 10-100 μm cracks that branch off the main front[24,25], begin to appear and coexist with steps. For v > $0.1c_R$, facets disappear altogether and successions of micro-branches, or branch-lines, dominate the surface.

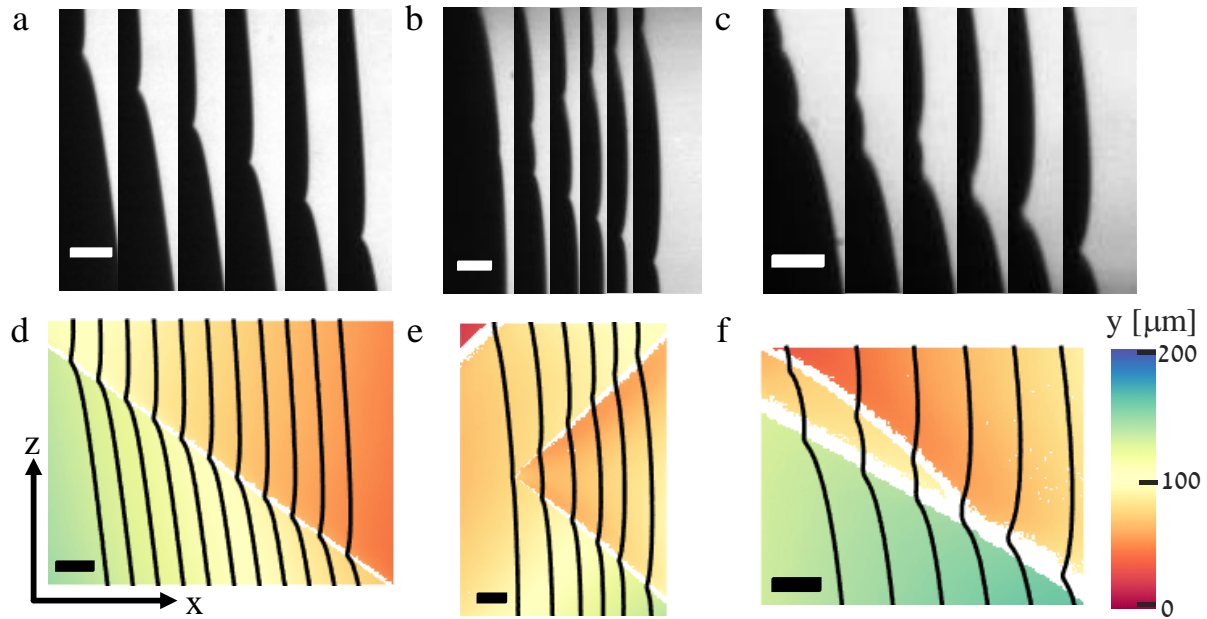

*Figure 2*. **Image sequences of facet-forming crack fronts and the resulting surface patterns**. Sequences of crack fronts (a) a single step at v= $0.005c_R$ displayed at $10ms$ intervals (b) two nucleating steps at v ~ $0.01c_R$ displayed at $5ms$ intervals and (c) merging step-lines at v ~ $0.01c_R$ displayed at $5ms$ intervals. (d-f) Profilometric measurements of the fracture surfaces created by the fronts depicted in (a-c). The fronts that formed them are overlaid in black. At steps, fronts are cusp-like with long-range curved tails. Scale bars are 200 μm long. White regions in the surface scan represent steep surface slopes.

During formation of facets, the in-plane (XZ) profile of the crack front is curved. We observe ~$50\mu m$ wide cusp-like regions at step locations from which long-range convex tails emanate, as shown in Fig. 2 where sequences of propagating fronts are superimposed on the fracture surface they created (see Extended Data Movies).

What determines step paths? Fracture surfaces yield a wide distribution of angles that step-lines form with the X axis. When step-lines exist simultaneously, their orientations can change during propagation (Fig. 3(a)). We define the step-line orientation as the angle $\theta$ relative to the local front normal, defined by the angle $\beta$ relative to X. While $\beta$ varies between $-40°$ to $+40°$ due to both long-range curvature, induced by coexisting step-lines, and global front tilts, $\theta$ is narrowly distributed; $\theta =$

$43° \pm 5°$ (Fig. 3(b)). In addition, we find that local crack velocities along the front vary in proportion to the local front slope, $\partial x/\partial z$ (Fig. 3(c)).

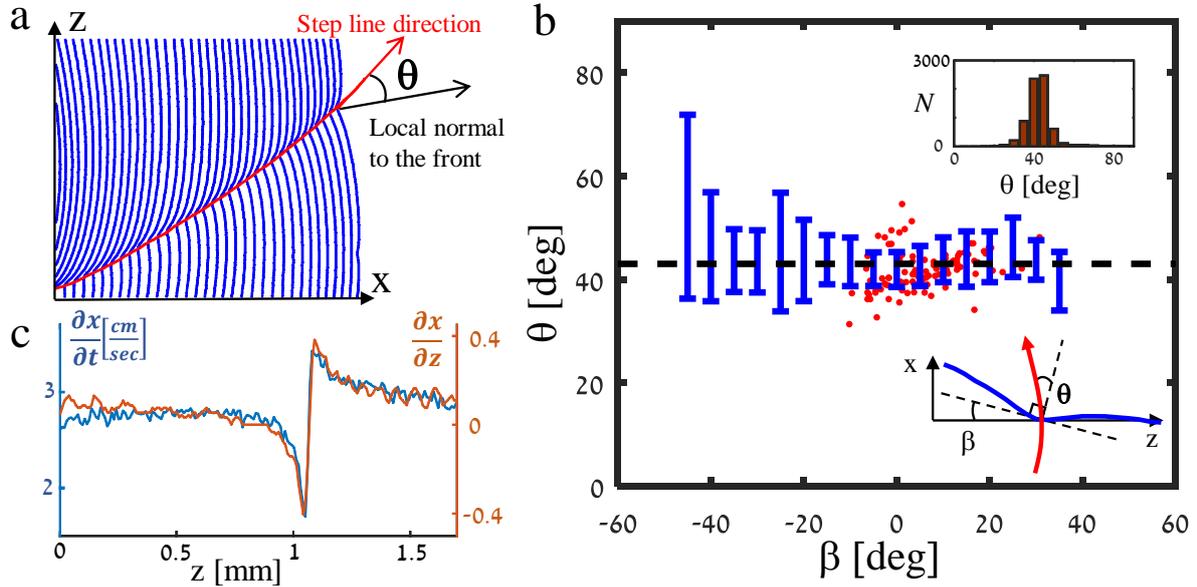

*Figure 3*. **Step-lines propagate at a constant angle to the local front normal in the crack frame**. (a) A sequence of crack fronts displayed at $0.27 ms$ intervals (in blue) form a step-line. The red line denotes cusp locations within the fronts associated with the step-line. As the step drifts along the front, its local orientation relative to the front changes, leading to step-line curvature. (b) The angle $-40° < \beta < 40°$ depicts the local front orientation (*bottom inset*).The angle $\theta = 43° \pm 5°$ between the step-line and the local front direction, remains constant. The blue bars show the range of $\theta$ in intervals of $5°$ in $\beta$, for $> 5000$ data points taken from 121 step-lines spanning $0.002 c_R < v < 0.05 c_R$. (*top inset*) Distribution of all $\theta$ values. Red dots are $\theta$ values corresponding to (a). (c) At each instant, local velocity $\partial x/\partial t$ varies in proportion to the local slope $\partial x/\partial z$, i.e. the crack front contains a "travelling wave" component $x(z,t) = az + vt + f(z - ut)$ where $u/v = tan(\beta \pm \theta)$ (the sign of $\theta$ is positive (negative) for downward (upward) propagating steps).

What determines front shapes? We now relate the *in-plane* front curvature to dissipation arising from the *out-of-plane* nature of the steps. Steps increase fracture surface area, hence locally increase dissipation. While this is a 3D effect, we will treat the front as a singular line confined to a planar fracture surface, since steps are confined to small regions. In this picture, the fracture energy varies as $\Gamma = \Gamma_0(v)(1 + \delta A(z))$, where $\Gamma_0(v)$ is the cost to create a unit surface area[16,26]. $\delta A(z)$ is the relative increase in surface area, which is strongly peaked at a step, but decays to zero elsewhere. The crack will propagate only if, locally, $G = \Gamma$ everywhere along the front. Since velocities are small, $\Gamma_0(v)$ is virtually constant, and any increase $\delta A$ must be compensated by a local increase of $G = G_0 + \delta G$, where $G_0 = \Gamma_0$. When $\delta G$ arises from a perturbation of a straight front $x(z) = x_0 + \delta x(z)$[13], $G = \Gamma$ yields

$$\frac{\delta G}{G_0} = -\frac{1}{\pi}\int \frac{dz'}{z-z'}\frac{\partial \delta x}{\partial z'} = \delta A(z) \ . \tag{1}$$

For a localized $\delta A$, Eq. (1) produces a locally concave profile with long-range logarithmic tails, as seen when interfacial crack fronts encounter a tough strip[27-29]. We find that front deformation due to a step at $z = 0$ is well-described by the asymmetric profile $\delta x(z) = cz + \frac{H}{\pi}\left[\frac{1}{2}\log\left(1+\frac{4z^2}{W^2}\right) \pm \alpha \arctan\left(\frac{2z}{W}\right)\right]$; where $H$ is the in-plane amplitude and $W$ is the localization scale. $\pm\alpha$ quantifies asymmetry that characterizes our front data. Fronts also exhibit a global slope $c$. Fig. 4(a) presents a typical example of >5000 crack fronts (taken from 25 isolated step-lines). The parameters $W = 50^{+10}_{-3}\mu m$ and $\alpha = 0.24 \pm 0.08$, are approximately constant in our data (see Methods, Extended Data Fig. 1). The asymmetry sign $\pm\alpha$ always coincides with the step drift direction along the Z axis: $+\alpha$ ($-\alpha$) when the step drifts in the positive (negative) Z direction, as shown in Fig. 4(b). Profilometric data reveal that step profiles are also asymmetric (see inset in Fig. 4(c)).

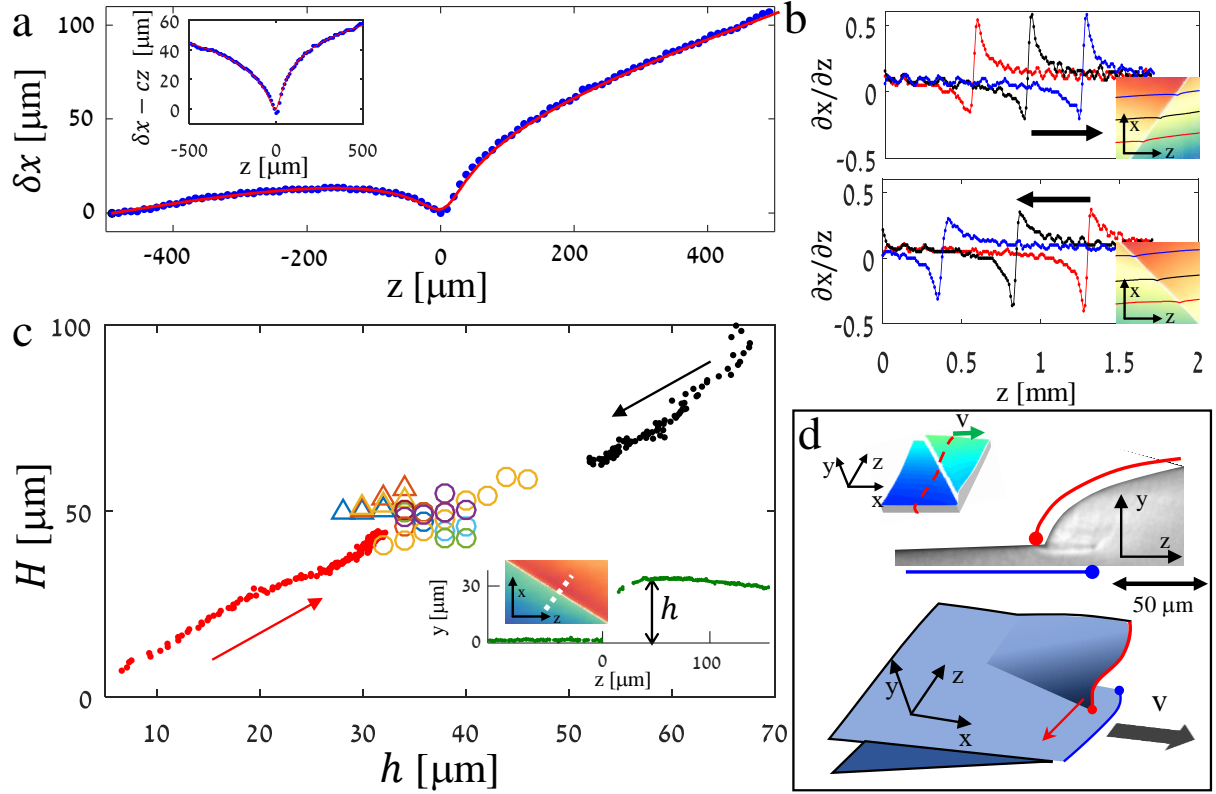

Figure 4 **Local dissipation at a step determines the long-range shape of the front**. (a) A typical front profile around a step. Measurements (blue dots) are in excellent correspondence to the functional form $\delta x(z) = cz + \frac{H}{\pi}\left[\frac{1}{2}\log\left(1+\frac{4z^2}{W^2}\right) \pm \alpha \arctan\left(\frac{2z}{W}\right)\right]$ (red line) of width $W = 50\mu m$ and asymmetry $\alpha = 0.24$. (*inset*) High resolution comparison after removing the global tilt $cz$. (b) Local

front slopes reveal distinct asymmetry. The asymmetry sign always agrees with step-line drift; drifts are positive (*top*) negative (*bottom*) for positive (negative) $\alpha$. Insets: fracture surfaces formed by overlaid fronts. (c) In-plane front amplitudes $H$ are proportional to out-of-plane step heights $h$; $H \sim 1.4\,h$. Red dots depict step-line rapid growth following the double nucleation event in Fig. 2(b,e). Black dots depict step-line height decay following a merging event similar to Fig. 2(c,f). Arrows depict growth (red) decay (black). Triangles and circles depict step-lines near steady-state, representing $H$ values averaged over step heights $h$ in $2\mu m$ intervals. Different colors represent single step-lines (triangles) and step-lines coexisting with other surface structures (circles). (*inset*) A typical step profile $y(z)$ along a section (dotted line) normal to the step-line presented. (d) The formation of a hidden surface by a discontinuous front. (*top*) A micrograph of the YZ section of a step. One segment of the step (red) curves and connects to the second flat segment (blue) which extends beneath the step to form a hidden surface. (*top inset*) Surface scan showing the cross-section orientation. (*bottom*) Topology of a facet-forming crack front. The curved segment (red) lags behind the flat segment (blue) while drifting along $z$ (red arrow).

The in-plane profile $\delta x(z)$ is equivalent, by means of Eq. (1), to the localized and asymmetric distribution $\delta A(z) = \frac{2H}{\pi W}\left(1 \pm \alpha \frac{2z}{W}\right)/\left(1 + \frac{4z^2}{W^2}\right)$; with $H = \int \delta A(z)\,dz$, the total surface area increase per unit crack length. In Fig. 4(c) we compare values of $H$, to the directly measured step heights $h$ (Methods). As steady-state values of $h$ are relatively constant ($h \sim 40\,\mu m$), a wide range of $h$ was obtained by considering two cases of growth after nucleation and decay following step merging. Remarkably, both in-plane amplitudes $H$ and out-of-plane step heights $h$ are proportional; $H \sim 1.4\,h$. Notably, $H > h$, while the surface area increase per unit crack length for a step of width $w$, $\sqrt{h^2 + w^2} - w$, which is always less than $h$, should ostensibly equal $H$.

The reason that $H > h$ becomes apparent upon a closer look at the step structure. Slicing the sample parallel to the YZ plane (Fig. 4(d)) reveals a hidden fracture surface that extends beneath the step. This hidden surface is part of a larger flat surface that comprises one side of the step (denoted with a blue line). The other side of the step (denoted in red) is *curved* and terminates abruptly at the point where it meets the flat surface. The slope of the step (inset of Fig. 4(c)) is steep near this point and becomes shallower as we move away from it. As noted in previous studies[6,7], these observations indicate that the crack front is separated into two discontinuous segments that overlap at a step to form a continuous fracture surface, with the curved segment lagging behind the flat segment. The overlap is responsible for the formation of the hidden surface. Correct estimation for the total increase in step surface area must include this surface. The relation $H \sim 1.4\,h$ is recovered if $w \sim h$, which is consistent with profilometric data (see Extended Data Fig. 2).

We have shown that a step is a localized energy sink that generates long-range elastic deformation along the crack front. The path of the step-line, however, is determined by a local criterion, i.e. it forms a constant $\theta \sim 43°$ angle with the local propagation direction. Once $\theta$ is selected by the system, the success of Eq. (1) indicates that local velocities $\partial x/\partial t$ are enslaved to front geometry, $\partial x/\partial z$, as seen in Fig. 3(b). A number of open questions remain.

Why are steps stable? Crack fronts that form steps are composed of two discontinuous overlapping segments, one necessarily lagging behind the other (Fig. 4(d)). Step stability can be clarified by considering two limiting cases. Linearly perturbing a *single* continuous front both in-plane ($\delta x$) and out-of-plane ($\delta y$) generates two decoupled stress contributions[10]. $\delta x$ induces tensile stresses that try to straighten the crack front, while $\delta y$ generates shear stresses that tend to flatten it. For two *overlapping* fracture planes the picture changes[30]. When a crack splits into two branches, induced shear stresses cause the two branches to "repel" each other; repulsion strength decreasing with branch separation. The repulsion is strong for small step heights and the two overlapping fracture surfaces are prevented from merging. On the other hand, the two segments of the crack front always retain coherence; macroscopically (Fig. 4(a)) the system behaves as a single in-plane crack subject to restoring tensile stresses. Plausibly, steps grow following nucleation due to branching repulsion and are later stabilized by the long-range restoring shear stresses that act along the front[10]. The non-trivial topology of step-forming crack fronts, then, prevents them from decaying to a flat state; crack faces cannot be continuously deformed to a flat configuration, the hallmark of a topological defect[6,7].

Why do steps drift along the front? Steps break reflection symmetry along the front ($\alpha \neq 0$ and $\theta \neq 0$). This suggests that stresses surrounding the step are distributed asymmetrically, which would lead to a bias in the step path. The value of $\theta$, we believe, cannot be determined by the planar theory. The selection of $\theta$ must arise from the 3D configuration of the discontinuity at the step. Revealing the nature of the step requires a more rigorous discussion of the complex mixed-mode stresses that surround it.

Given the importance of three dimensionality to step stability, direction and dissipation, one might wonder why the planar theory works at all? Here, again, we draw an analogy from systems with topological defects. In these systems small-strain elasticity breaks down within a localized core region. In the case of an edge dislocation this core is the size of a few atomic distances. Outside of this region, elasticity remains valid, since strains are small. Similarly, outside of step regions front deformations are still very much governed by in-plane elastic tension, although crack front segments lie in planes separated by tens of microns.

To conclude, fracture steps provide a fundamental way to break crack front continuity. Our direct observations of crack front propagation during step formation provide a path-selection criterion for step-lines, while the linear relation between in-plane and out-of-plane deformations provides a close link between 3D instability and planar fracture mechanics. These basic observations provide an empirical basis for a fundamental theory of 3D crack front dynamics.

**METHODS SUMMARY**

The supplementary Methods section contains details of gel preparation, fracture procedures, surface profilometry measurements, image analysis and crack front characterization. Two supplementary figures support statements that are made in the main text: the first showing the statistics of the parameters $W$ and $\alpha$, and the second depicting the proportional variation of step width $w$ with step height $h$. Finally, we provide movies of crack front propagation for the three cases shown in Fig. 2.

*Acknowledgements*

J.F. and I.K. acknowledge the support of the Israel Science Foundation (grant no.1523/15), as well as the US-Israel Bi-national Science Foundation (grant no. 2016950). I.K. thanks I. Svetlizky and E. Katzav for fruitful discussions about step stability. I.K. is grateful to P. M. Chaikin for an enlightening conversation on the complexity of fracture surfaces.

*Author Contributions*

I.K. and G.C. designed the gel preparation method and fracture experiments. I.K. synthesized the gel samples, performed the fracture experiments and surface profilometry and analyzed data. J.F. conceived the 3D crack front imaging, initiated and supervised the research. The manuscript was written by all authors.

# Supplementary material to "Topological defects govern crack front motion and facet formation on broken surfaces"

## *Methods*

**Gel preparation**. We prepared 14% (w/v) polyacrylamide gels cross-linked with 2.6\% (w/w) N,N'-Methylenebisacrylamide. Bulk polymerization was initiated with ammonium persulfate and catalyzed with TEMED. Solution was then poured into a home-made mold. The mold was constructed of two optically flat glass bars that are placed parallel to each other on a glass base plate. The faces of the resulting gel, which inherit the flatness and parallelity of these bars, enabled undistorted optical access through the two XZ faces of the sample. A thin acrylic plate is fitted between the bars to act as a cover, and a set of machined acrylic spacers supported the cover above the base plate leaving the two YZ sides open throughout the polymerization process. During polymerization, the mold was surrounded by a "bath" of the polymer solution. Prior to polymerization, the mold parts and bath were rinsed with soap, dried and then cleansed with ethanol. After assembling the mold, the solution was poured so as to fill the bath and submerge the mold. This technique prevents the sample from having free surfaces that can create anisotropic stresses during polymerization and destroy the optical uniformity of the gel. In order to prevent polymerization inhibition by atmospheric oxygen, the bath together with mold and solution were placed in a sealed container filled with argon gas. Polymerization was completed within 90 minutes, after which the gel sample was carefully extracted from the mold and cut to a rectangular shape of dimensions $54 \times 94 \times 4.6 \ mm^3$ $(X \times Y \times Z)$.

**Fracture experiments**. The sample was loaded within two grips positioned at the ends of the sample's Y dimension leaving $56 \ mm$ of free material. Fractures were initiated by two different methods. When very slow fracture was required, a small notch was imposed at the sample's edge at its mid-plane and an initial seed crack was made, using a scalpel at the notch's center. This crack then propagated into the sample by applying opposing point loads to the notch faces until a sharp seed crack of desired length was obtained. Only then was tension applied to the sample (by displacing the grips at a rate of $50 \mu m/sec$) until slow (e.g. v= $0.005 c_R$) crack propagation initiated. A second method was used to obtain more rapid fracture. In this method, prior to fracture, we first applied a constant displacement to the sample by translating both grips anti-symmetrically so that the center XZ plane was stationary. A seed crack was then introduced by pushing a glass fiber of diameter $100 \mu m$ via a translation stage through the edge of the sample at its mid-plane (in Y) until propagation initiated. Using this method, the sample was slightly overstressed prior to initiation of propagation and the more rapid velocities were obtained. In both methods, these careful crack initiation procedures were necessary to ensure that no initial surface structure was induced prior to fracture initiation.  Images of the moving front were obtained by shining a beam of collimated (LED) light through the bulk of the gel in the Y direction. The grips had built-in optical windows that enabled collimated illumination and imaging of the fracture plane. A 270mm lens collected the outgoing light and projected the image of the XZ mid-plane of the sample onto the sensor of a high-speed camera (Y4-S2, IDT Inc.). We recorded images at a spatial resolution of $10\ \mu m$ per pixel and at rates of $10,000 - 15,000$ frames per second. Due to boundary effects, we had a clear view of only the center $2\ mm$ of the total $4.6\ mm$ sample thickness. During crack propagation through the gel, the high curvature at the crack front deflected the incoming light. As a result crack fronts appeared as a sharply defined shadow that progressed into an otherwise uniformly illuminated frame.

**Surface profilometry.** Immediately following each experiment, we made a cast of the fracture surface using polyvinyl siloxane (Elite HD+, Super Light Body, Zhermack). The cast captures surface patterns in microscopic detail. This was verified by comparing height measurements of a $120\ \mu m$ grid lithography etched on glass with a cast of the same grid. The surface casts were measured with an optical profilometer (Contour GT-I, Bruker Co.) to produce 3D height maps with a lateral resolution of $2\ \mu m$ and a vertical resolution of $1\ \mu m$.

**Image processing.** Crack fronts were extracted from shadow images at sub-pixel ($\sim 3\mu m$) resolution. Shadow images were divided by a background image and the front position was defined using a threshold of 0.5. To find the point where relative image intensity passed the threshold we cubically interpolated image intensity along the x axis, enabling our sub-pixel resolution.

**Matching crack fronts and fracture surface patterns.** Fracture steps produced a cusp-like deformation in the crack front. We first detected the cusp locations at each front and obtained the step-line pattern in the crack frame, which was elastically deformed relative to the material rest frame. We then extracted the contours of the step-lines in the material rest frame from the profilometric measurement. The two patterns were then superposed by imposing both a global translation and homogenous deformation.

**Estimation of $W$ and $\alpha$ for isolated step-lines.** We estimated $W$ by numerically differentiating the crack front (using a Savitzky-Golay filter of order 2 and with a window size of 5 pixels), and measuring the distance between the two extrema that appear at the step location. For the data in Fig. 4(b) $W = 50^{+10}_{-3}\mu m$ is constant regardless of the variation of $H$ (see Extended Data Fig. 1). We therefore fixed $W = 50\mu m$ and initially performed fits over all our data with $H$ and $\alpha$ as free parameters. We found that the value of $\alpha = 0.24 \pm 0.08$ was uncorrelated with $H$. The value of $\alpha$ justifies our estimation of $W$; the analytical derivative of the functional form $\delta x(z) = cz + \frac{H}{\pi}\left[\frac{1}{2}\log\left(1 + \frac{4z^2}{W^2}\right) \pm \alpha \arctan\left(\frac{2z}{W}\right)\right]$, i.e. $\frac{\partial x(z)}{\partial z} = c + \frac{2H}{\pi W}\frac{\frac{2z}{W} \pm \alpha}{1 + \frac{4z^2}{W^2}}$ has extrema that are separated approximately by $W$ with an error term $O(\alpha^2)$.

Since $\alpha^2 \ll 1$ the numerical estimation of $W$ is correct. The values of $H$ in Fig. 4(b) were in the end obtained by fixing both $\alpha$ and $W$ and fitting to $\delta x(z)$ with $H$ as the sole free parameter. In cases where more than one step existed along the front at a given time, it was necessary to add a global quadratic term to $\delta x(z)$ to account for the long-range curvature induced by neighboring steps.

**Fits to double nucleation (e.g. Fig. 2(b,d)).** In order to extract the amplitude $H$ in double nucleation cases, we used a superposition $\delta x(z - z_1) + \delta x(z - z_2)$ where $z = z_{1,2}$ are the locations of the two steps. As we observe that both steps grow in exactly the same way, we used the same amplitude $H$ for both $\delta x(z - z_1)$ and $\delta x(z - z_2)$.

## *Extended Data*

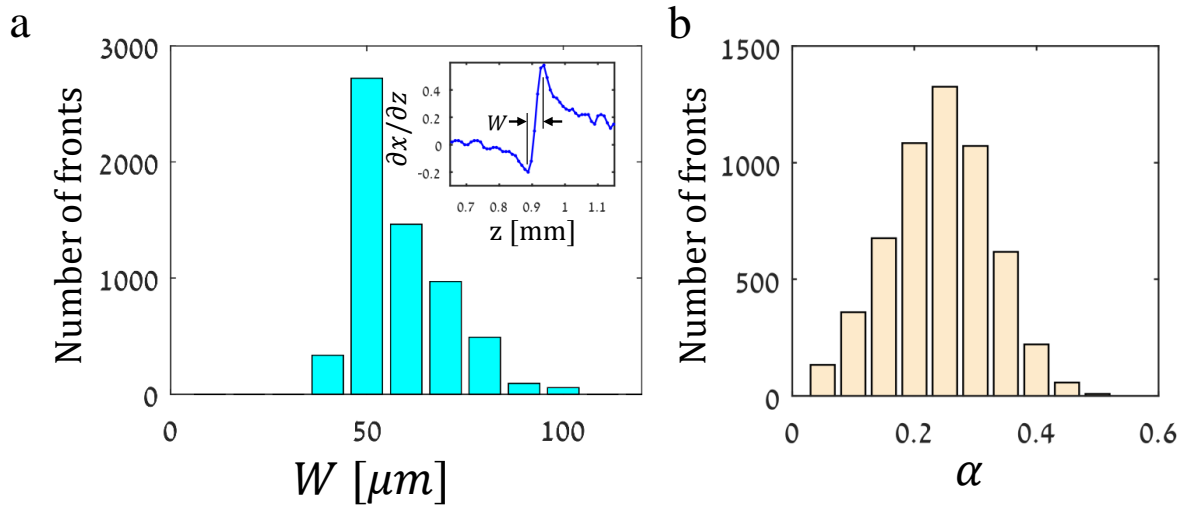

*Extended Data Figure 1.* **The parameters $W$ and $\alpha$ extracted from isolated step-lines.** (a) Values of $W$ are concentrated around $W = 50^{+10}_{-3} \mu m$. (*inset*) $W$ is determined by measuring the distance between the minimum and the maximum of the front derivative $\partial x / \partial z$. (b) $\alpha$ (asymmetry) values determined from front profiles by fitting to $\delta x(z) = cz + \frac{H}{\pi} \left[ \frac{1}{2} \log \left( 1 + \frac{4z^2}{W^2} \right) \pm \alpha \arctan \left( \frac{2z}{W} \right) \right]$ assuming a constant $W = 50 \mu m$. The distribution yields $\alpha = 0.24 \pm 0.08$

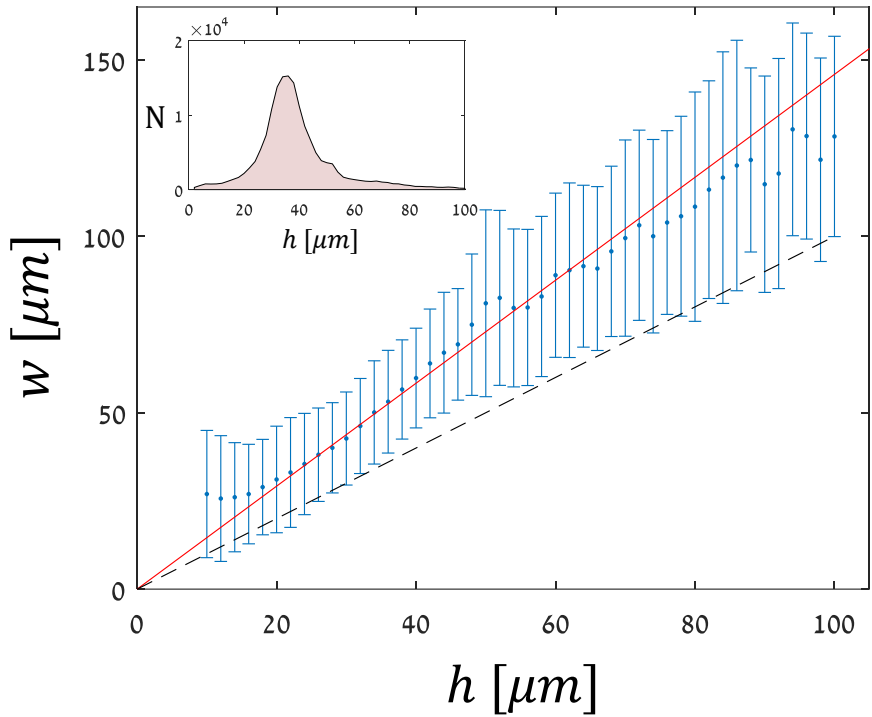

*Extended Data Figure 2.* **Step width $w$ grows with step height $h$.** Blue bars show the range of $w$ measured by profilometry when averaged over $2\mu m$ intervals of $h$. Linear regression (red line) yields a proportionality coefficient of $1.4 \pm 0.4$. The dashed black line, $w = h$, is one standard deviation below the mean. (*inset*) Number of data points within each $2\mu m$ interval of $h$. A total of ~155,000 points was considered.

*Extended Data Movie 1.* Propagation of a crack front forming a single step-line at steady state over $50ms$ (interval between frames is $0.5ms$) (see Fig. 2(a,d)).

*Extended Data Movie 2.* Propagation of a crack front throughout the simultaneous nucleation and growth of two step lines over $30ms$ (interval between frames is $0.3ms$) (see Fig. 2(b,e)).

*Extended Data Movie 2.* Propagation of a crack front during the merging of two step-lines over $25ms$ (interval between frames is $0.3ms$) (see Fig. 2(c,f)).